# Optical fiber for remote sensing with high spatial resolution

Michael Eiselt, Florian Azendorf, André Sandmann, ADVA SE, Meiningen, Germany

## Abstract

The use of optical fiber as sensor as well as transmission medium for sensing data is discussed, enabling the use of optically active sensors without power supply at distances of tens of kilometers. Depending on the interrogation system, a spatial resolution of less than a millimeter can be achieved. The basic sensing principle is optical time-domain reflectometry (OTDR) with direct detection or coherent detection of the Rayleigh back-scattered or Fresnel reflected signal. Spatial resolution is improved by a cross-correlation between the transmitted sequence and the received signals.

## 1 Introduction

The use of transmission fibers as optical sensor has recently gained strong scientific attraction. Refractive index and polarization changes in the fiber are monitored to obtain information about environmental parameters in the vicinity of the fiber, including temperature, vibrations, or stress to the fiber. The application is two-fold. On one hand, any impact on the fiber infrastructure can be detected early, protecting the optical transmission system [1]. In addition, environmental events, like submarine earthquakes, can be detected [2].

The sensing efficiency of the fiber can be improved by writing Bragg gratings into the fiber, which change their reflection properties depending on environmental impacts. The refractive index and polarization changes can be interrogated remotely, using the same optical fiber to transport the optical probe signal to and from the sensor fiber section. No power supply is required at the location of the sensor.

In this paper, various methods are discussed to utilize the fiber as a sensing medium. With appropriate probe signals, a spatial resolution of less than a millimeter can be obtained [3], even over fiber lengths of several kilometers.

## 2 Optical time domain reflectometry with direct detection

Environmental effects can change a variety of optical fiber parameters. It is well known that temperature and strain can shift the Brillouin frequency, while the Raman frequency is sensitive to temperature variations in glass fibers [4]. In this paper, we will restrict the consideration to the change of the fiber refractive index, induced also by temperature and strain variations, and leading to a change in propagation time and phase of the propagating optical signal. Furthermore, in this paper we will only consider the effects of Rayleigh scattering and Fresnel reflections, both leading to returning parts of a signal propagating along a fiber.

In a basic setup, as shown in **Figure 1,** to sense the variations along a fiber, an optical pulse is injected into a fiber. Scattering and reflections along the fiber will lead to part of the signal being returned to the fiber input, where the signal is detected. Based on the time difference between the injected pulse and the returned signal, the location of the reflection can be determined. In standard applications of this optical time-domain reflectometry (OTDR) a pulse length on the order of tens of nanoseconds to a few microseconds is used, resulting in a spatial resolution of the reflection location of a few meters to kilometers, based on the typical round-trip time (RTT) of the optical signal in standard telecommunication fiber of 10 ns per meter. Recently, instead of a single long optical pulse, the use of pulse sequences at a data rate of 10 Gbps was reported [5], yielding, with over-sampling and signal processing, to a temporal resolution of less than 10 ps, which is the RTT of 1 mm fiber. A cross-correlation between the returned signal and the transmitted bit sequence improves the signal-to-noise ratio, such that the use of a 1000-bit sequence would lead to a similar sensitivity as a 100-ns long single pulse.

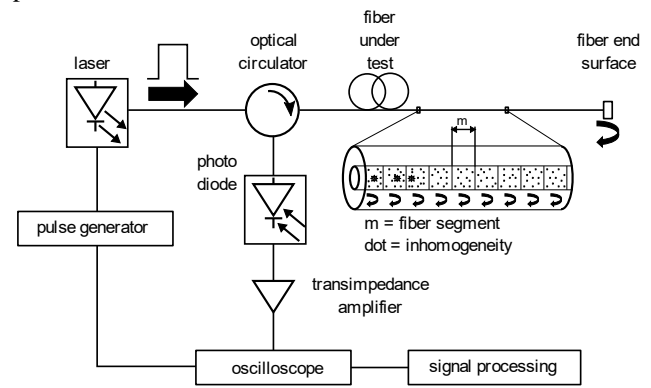

**Figure 1** Basic setup for OTDR fiber sensing

An application of this high temporal resolution is the temperature measurement in a long fiber. With a typical temperature coefficient of the propagation time in standard telecommunication fiber of $7\times10^{-6}$ per Kelvin, the RTT of one kilometer of fiber changes by 70 ps per Kelvin.

**Figure 2** shows the variations of the RTT through 8.5 km of fiber, deployed about 1 meter deep in the ground, over a time period of two weeks. In addition, the temperature variations over this time period, as obtained from a nearby weather station, are shown. It can be seen that the RTT variations closely track the temperature evolution, considering that the temperature 1 meter in the ground is low-pass filtered with a time-coefficient of approximately 12 days [6].

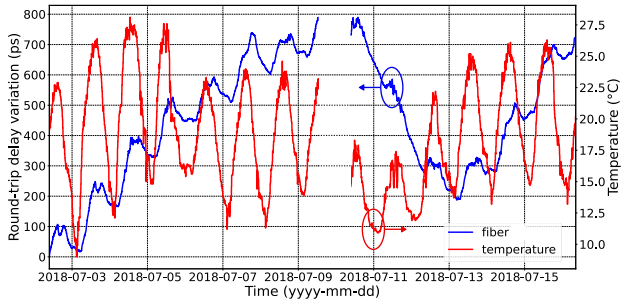

**Figure 2** Blue: variations of the round-trip time of 8.5 km fiber over two weeks. Red: temperature variations over this time, obtained from a weather station.

## 3   Coherent signal detection

While the results in the previous chapter were obtained with direct detection of the returned optical signal on a photo diode, coherent detection (CD) promises a two-fold improvement. On one hand, the received signal variations are amplified by the superposition of the signal with the local oscillator. On the other hand, CD allows the detection not only of the signal intensity, but also of the phase and the polarization of the signal, and thus yields further information on the sensor fiber. It also enables the use of phase encoding the transmitted bit sequence and thus to further improve the receiver sensitivity.

The setup of a coherent OTDR differs from the direct detect variant only in the replacement of the receiver photo diode by a coherent receiver with four electrical output signals, corresponding to the in-phase and quadrature components of the signal in both polarizations. The local oscillator of the receiver is split off the transmitter laser before modulation and is connected to the respective port of the coherent receiver.

To demonstrate the improved sensitivity of the coherent detection, **Figure 3** shows the reflected and backscattered signal from a 60-m fiber with a number of different connector types. Reflections from the connectors can be noted with different intensities. Calibrating the reflection amplitude to an open connector with 14 dB return loss, the reflection coefficients in **Figure 3** can be identified as between -51.3 dB and -63.3 dB. In addition, backscattering from the fiber between the connectors can be observed slightly above the noise level of –70 dB. This fits well to the expected backscattering level of approximately -89 dB for the probe pulse width of 200 ps.

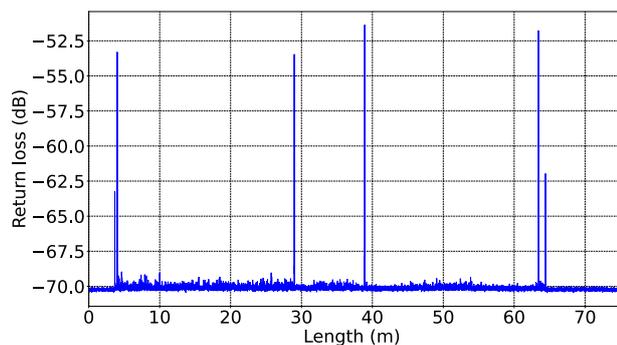

**Figure 3** OTDR trace of 60 m fiber, obtained with coherent detection with a probe bit rate of 5 Gbps.

## 4   Interrogation of fiber Bragg gratings

The high temporal resolution of the coherent correlation OTDR allows the interrogation of closely spaced Bragg gratings in the fiber. These fiber Bragg gratings (FBGs) reflect a certain wavelength range, depending on the grating period. As the period depends on the refractive index variations, the Bragg wavelength changes with temperature with a rate of typically 0.01 nm/K.

A 100-m long fiber was inscribed with 2000 weakly reflecting FBGs (-30 dB per grating), separated by 50 mm. A coherent correlation OTDR with a tuneable optical frequency was used to measure the reflection spectra for all FBGs, varying the optical frequency in steps of 1 GHz. **Figure 4** shows the reflection amplitudes for the first 200 gratings over a frequency range from 193.365 THz to 193.41 THz (corresponding to wavelengths from 1550.436 to 1550.075 nm). A variation of the reflection wavelength with a period of approximately 10 FBGs can be observed, corresponding to the fiber spool circumference and caused by strain variations on the spool [7].

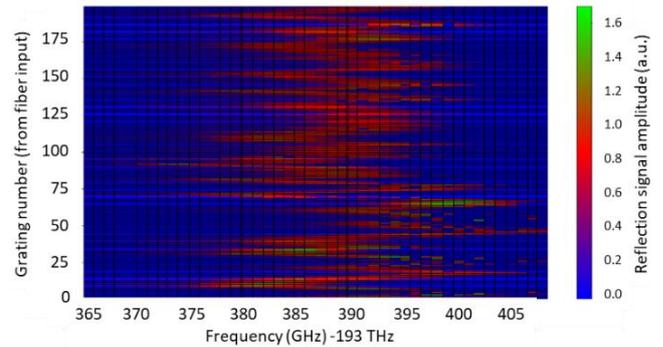

**Figure 4** Reflection spectra from 200 FBGs along a fiber, spaced by 50 mm, showing variations of the center wavelength with a period of ~0.5 meter.

## 5   Phase sensitive applications

As mentioned before, coherent detection of the backscattered signal yields information about the round-trip phase of the probe signal. While for a static fiber with constant external impacts the round-trip phase stays constant over time and the receiver phase is constant (except for the effect of laser phase noise), a change in the fiber refractive index at one location leads to a phase change of all probe signals passing this location during the fiber round-trip. This enables a spatial resolved monitoring of even small phase changes along the fiber, stemming from strain, temperature variations, or vibrations. This method can even be used to monitor acoustic vibrations from sounds in the fiber vicinity [8].

While the direct detection OTDR is able to monitor small temperature changes of a long fiber section based on the RTT variation, a coherent correlation OTDR measurement with a high spatial resolution enables the monitoring of small temperature variations over fiber lengths of a few millimeters. Given a temperature dependence of the refractive index in standard fiber of $10^{-5}$/K, the round-trip phase over 10 mm fiber at a wavelength of 1550 nm changes with temperature by approximately 45 deg/K. This enables a small temperature sensor with a sub-Kelvin temperature resolution and a low thermal capacity.

# 6  Summary

The refractive index of optical fiber is a parameter that reacts to a variety of external parameters, foremost temperature, strain, and vibrations. Therefore, optical fiber can be used as a distributed sensor. Enhancing the fiber structure, e.g. by writing a Bragg grating, can improve the sensitivity. The same fiber can also be used to transmit the interrogating light signal to the sensor location and back, enabling to place the sensor at remote locations in a fully passive mode.
It has been shown that different types of the interrogator are possible, all of them based on optical time-domain reflectometry and using correlation methods to improve the spatial resolution.